# Energy Distribution of Nanoflares in Three-Dimensional Simulations of Coronal Heating

C. S. Ng[a] and L. Lin[b]

[a]*Geophysical Institute, University of Alaska Fairbanks, Fairbanks, Alaska 99775, USA*
[b]*Space Science Center, University of New Hampshire, Durham, NH 03824, USA*

**Abstract.** In a recent computational campaign [Ng *et al.*, Astrophys. J. **747**, 109, 2012] to investigate a three-dimensional model of coronal heating using reduced magnetohydrodynamics (RMHD), we have obtained scaling results of heating rate versus Lundquist number based on a series of runs in which random photospheric motions are imposed for hundreds to thousands of Alfvén time in order to obtain converged statistical values. Using this collection of numerical data, we have performed additional statistical analysis related to the formation of current sheets and heating events, or nanoflares [Parker, Astrophys. J. **330**, 474, 1988]. While there have been many observations of the energy distribution of solar flares, there have not been many results based on large-scale three-dimensional direct simulations due to obvious numerical difficulties. We will present energy distributions and other statistics based on our simulations, calculated using a method employed in [Dmitruk & Gómez, Astrophys. J., **484**, L83, 1997]. We will also make comparisons of our results with observations.

## INTRODUCTION

As is well known to the space and astrophysics community, the coronal heating problem, i.e. how the solar corona can be heated to millions degrees while the photosphere is only about 6000 K, remains a major unresolved problem. One possible mechanism, among many others, was proposed by Parker [1] that the heating is due to the formation of current sheets (tangential discontinuities) when magnetic footpoints at the photosphere are twisted randomly for a sufficiently long time such that quasi-equilibrium of the magnetic structures can no longer be maintained. Since then, this mechanism has been studied theoretically and numerically by many papers, too many to be listed here. More extensive references can be found in our recent paper [2], in which we revisited this model using three-dimensional 3D reduced magneto-hydrodynamics (RMHD) numerical simulations. In that paper, we have presented long-time simulations of the random photospheric twisting process over hundreds to thousands of Alfvèn times (time for Alfven wave to propagate along a coronal loop from one end to the other) so as to obtain accurate average heating rate $W$, for different levels of resistivity $\eta$ (i.e. ~ inverse of the Lundquist number $S$). Based on these numerical results, we concluded that $W$ tends to a level that is independent of $S$, while the production of the magnetic energy depends only weakly on $S$, in the high $S$ limit. That paper also gave theoretical arguments based on the random walk nature of photospheric footpoint motions to justify why the scalings differ in the high $S$ limit from those of a similar study done many years ago [3] which was limited to a lower $S$ regime given prevailing computational power. Our series of long-time simulations has resulted in the accumulation of a large volume of numerical data, which can be further investigated for other statistical properties. Some of these are the subjects of this paper, i.e., statistical distributions of heating events, which can model the so-called "nanoflare" heating concept [4].

In Ref. [4], Parker argued that most contribution to coronal heating is from many small heating events, or nanoflares, below the level of microflares (~$10^{27}$ erg), that are formed by random twisting of magnetic footpoints. A simple criterion for the importance of nanoflare heating can be obtained in a straightforward manner if the energy distribution has a power-law dependence on energy, i.e., $dN/dE \propto E^{-\alpha_E}$ over the full range of flare energy $[E_{min}, E_{max}]$, where $dN$ is the number of flares with energy within the range $[E, E + dE]$ over a certain observation period. It is clear that if $\alpha_E > 2$, the total heating is indeed contributed more by flares with energy closer to (in logarithmic scale) the lower limit $E_{min}$. On the contrary, if $\alpha_E < 2$, most contribution is instead from flares with energy closer to the upper limit $E_{max}$ [5].

Ever since the nanoflare heating model by Parker, there have been many studies to try to test this criterion by comparing with observed flare distributions. These efforts are intrinsically limited by first the inability to resolve flares with low energy, and secondly by the small statistics of events with high energy. Thus, observed distributions are not reliable for the full range of $[E_{min}, E_{max}]$ but instead are just for a certain middle sub-range of it. Nevertheless, these observations still give important constrains on the physical picture of nanoflare heating. One of the early results did report a power-law index of $\alpha_E \sim 1.8$ [5], which raised doubt over the importance of nanoflare heating mechanism. Since then, there have been many more studies but most reported indices less than 2, as summarized in a recent review [6].

There have also been theoretical and modeling studies on the nanoflare heating mechanism, including dynamical models based on the concept of self-organized criticality (or avalanche models) such as Ref. [7], which reported an index of $\alpha_E \sim 1.4$. However, there have been fewer direct numerical simulations based on fluid equations. One such study did report an index of $\alpha_E \sim 1.5$ based on 2D MHD simulations [8,9]. In this paper, we will apply the method of analysis developed in Refs. [8,9] to analyze outputs from our 3D RMHD simulations and obtain statistical properties, such as energy distributions and power-law indices.

## NUMERICAL SETUP AND RESULTS

Following the usual setup for the Parker's model of coronal heating, our simulation box is a rectangular volume representing a straightened coronal loop, with a uniform background magnetic field along the $z$-direction. Due to the low beta situation in the corona, we simulate the dynamics using RMHD equations, which are as follows in dimensionless form:

$$\frac{\partial \Omega}{\partial t} + [\phi, \Omega] = \frac{\partial J}{\partial z} + [A, J] + \nu \nabla_\perp^2 \Omega \, , \qquad \frac{\partial A}{\partial t} + [\phi, A] = \frac{\partial \phi}{\partial z} + \eta \nabla_\perp^2 A \, , \qquad (1)$$

where $A$ is the flux function so that the magnetic field is $\mathbf{B} = \hat{\mathbf{z}} + \mathbf{B}_\perp = \hat{\mathbf{z}} + \nabla_\perp A \times \hat{\mathbf{z}}$; $\phi$ is the stream function so that the flow velocity is $\mathbf{v} = \nabla_\perp \phi \times \hat{\mathbf{z}}$; $\Omega = -\nabla_\perp^2 \phi$ is the $z$-component of the vorticity; $J = -\nabla_\perp^2 A$ is the $z$-component of the current density; $[\phi, A] \equiv (\partial \phi / \partial y)(\partial A / \partial x) - (\partial \phi / \partial x)(\partial A / \partial y)$; $\eta$ is the resistivity (inverse of the Lundquist number $S$); and $\nu$ is the viscosity (inverse of the Reynolds number). The normalization in Eq. (1) is such that magnetic field is in the unit of $B_z$ (which is a constant in RMHD); velocity is in the unit of $v_A = B_z / \sqrt{4\pi\rho}$ with $\rho$ being the density; length is in the unit of the transverse length scale $l$; the unit of time is $l/v_A$; $\eta$ is in the unit of $4\pi v_A l / c^2$; and $\nu$ is in the unit of $\rho v_A l$. The surfaces at $z = 0$ and $z = L$ (which is chosen to be unity in runs reported here) represent photospheric boundary with random flow velocity imposed to twist magnetic field lines based on the line-tied boundary condition. Double-periodic boundary condition is imposed in the $x$-$y$ directions and thus enables the application of standard pseudo-spectral method to increase accuracy in resolving fine structures in the $x$-$y$. Leapfrog finite difference method is used in the $z$-direction. To run the 3D code in higher resolutions, simulations are performed on parallel computers using MPI (Message Passing Interface). More recently, to obtain even longer runs at high resolutions, we have also modified the code and run it on machines with GPUs (Graphics Processing Units) using Nvidia's Compute Unified Device Architecture (CUDA) [10].

As random boundary flows twist the magnetic field, current sheets form and dissipate. Fig. 1(a) shows a few iso-surfaces of $J$ at an instance within a run. Following the method employed in Refs. [8,9], the identification of heating events is based on the output time series of the total heating rate $W = W_\eta + W_\nu$, where $W_\eta = \eta \int J^2 d^3 x$, $W_\nu = \nu \int \Omega^2 d^3 x$ with $W_\eta \gg W_\nu$ usually due to slow driving flow velocity on boundaries. Fig. 1(b) plots $W_\eta$ for two runs, showing the intermittent nature due to the random forming and dissipating of current sheets.

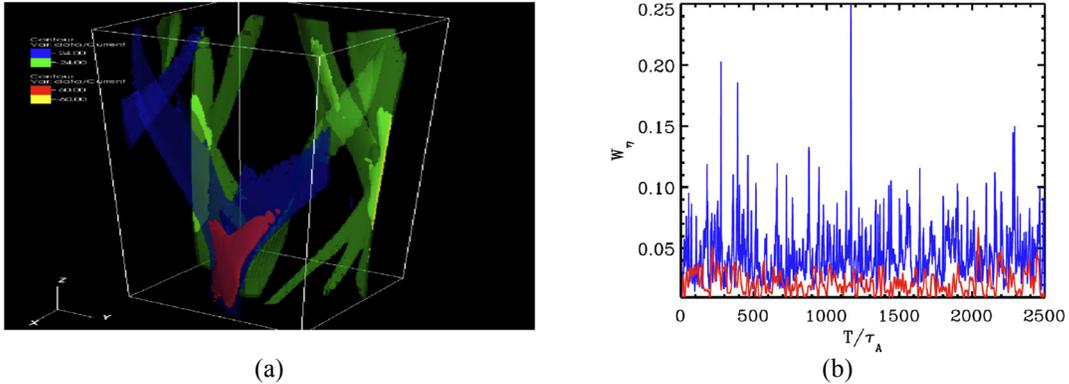

(a)          (b)

**FIGURE 1.** (a) 3D iso-surfaces of $J = -60, -24, 24, 60$ at a time taken from the run with $\eta = \nu = 0.000625$; (b) Time series of $W_\eta$ for the run with $\eta = \nu = 0.0003125$ (blue) and $\eta = \nu = 0.005$ (red).

The identification method proceeds with the subtraction of the average value of $W$ over time, or $\overline{W}$, and considers only the positive part. A portion in time during which $W - \overline{W}$ is all positive is then identified as a heating event. The duration of this event in time is denoted as $\tau$. The energy dissipated during this event, $E$, is defined as the time integration of $W - \overline{W}$ over this duration $\tau$. The maximum of $W - \overline{W}$ within this event is called the peak power $P$. The duration between an event to the next one is called the wait time $w$. Once events are identified, frequency plots of number of events can be made with respect to $E$, $\tau$, $P$, or $w$, as shown in Fig. 2 for one run.

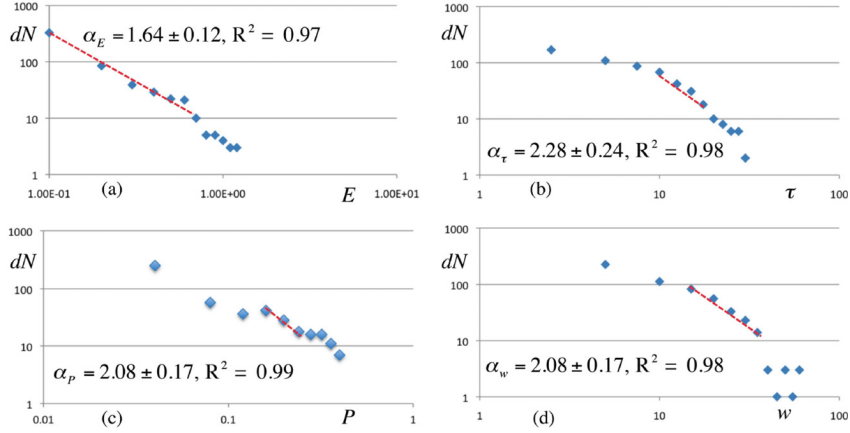

**FIGURE 2.** Frequency plot of number of events per bin ($dN$) versus (a) energy $E$, (b) duration $\tau$, (c) peak power $P$, (d) wait time $w$, for a case with $\eta = \nu = 0.0003125$ using a resolution of $256^2 \times 32$. Red dashed lines indicate the portion of data used for least square fit to obtain power-law indices $\alpha_E$, $\alpha_\tau$, $\alpha_P$, $\alpha_w$ with values shown on the graphs, along with the coefficient of determination $R^2$.

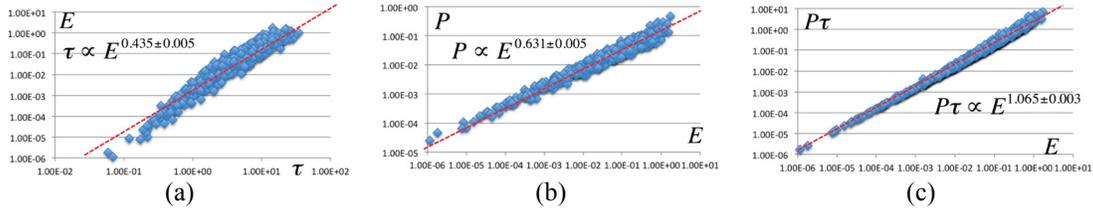

**FIGURE 3.** Scatter plot of (a) $E$ vs. $\tau$, (b) $P$ vs. $E$, (c) $P\tau$ vs. $E$ for a case with $\eta = \nu = 0.0003125$ using a resolution of $256^2 \times 32$. Red dashed lines indicate the best least square fit in log-log scales to obtain power-law relationship between quantities as indicated on the graphs.

As can be seen from Fig. 2, the data of each plot do not follow a single power law over the whole range since they do not lie on a straight line on the log-log plot. This is due to poor statistics near the high-value end because the number of events becomes small (less than 10), and the finite bin effect near the low-value end since the first bin might contain events over a range of a few decades of small values (or it might reflect a true deviation from a single power law). Therefore, to get a power law index, e.g. $\alpha_E$ such that $dN \propto E^{-\alpha_E}$ etc, only a portion of the data in the middle range which can fit better to a straight line (usually for $dN > 10$), as indicated by a red dashed line is used for least square fit. There is likely some uncertainty in choosing the straight-line portions. On the other hand, scatter plots between $E$, $\tau$, $P$, and $w$ can also be made to reveal possible correlations among them. As it turns out, some of these relations can be used to double-check some of the values of these indices obtained using the above process. In Fig. 3, scatter plots of $E$ vs. $\tau$, $P$ vs. $E$, and $P\tau$ vs. $E$ are shown for the same set of events. It is clear that all three plots indicate very good power law relations, especially the $P\tau$ vs. $E$ plot with a $P\tau \sim E^{1.065}$ relation, which is consistent with the physical meaning that the total energy of an event is roughly given by the peak power times the duration. Note that similar relationship, with $P\tau \sim E^{1.18}$ instead, was found in observations and was the reasoning behind the suggestion to use $P\tau$ as a proxy for $E$ since the former is much easier to measure than the latter [11]. Also, based on the fact that $P$, $\tau$ and $E$ are well correlated with each other, assuming that

$$\frac{dN}{dP} \propto P^{-\alpha_P}, \quad \frac{dN}{d\tau} \propto \tau^{-\alpha_\tau}, \quad \frac{dN}{dE} \propto E^{-\alpha_E}, \tag{2}$$

one can deduce the power law relation of $P\tau \sim E^\gamma$, with $\gamma$ given by

$$\gamma = \frac{(\alpha_E - 1)(\alpha_P + \alpha_\tau - 2)}{(\alpha_P - 1)(\alpha_\tau - 1)}. \tag{3}$$

Using $\alpha_P = 2.08$, $\alpha_\tau = 2.28$, and $\alpha_E = 1.64$ observed from Fig. 2, we get $\gamma = 1.09$, which is very close to the relation found from Fig. 3(c) with $\gamma = 1.065$. This provides a self-consistent check for the values obtained for the indices found in Fig. 2.

The same analysis was applied to outputs from three other runs with larger $\eta$. These require lower resolution and thus have longer run time and larger number of heating events. The frequency plots for the case with the largest $\eta$ are shown in Fig. 4.

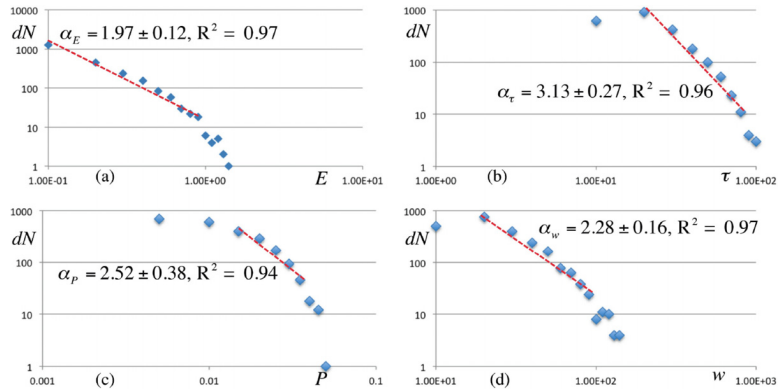

**FIGURE 4.** Frequency plot of number of events per bin ($dN$) versus (a) energy $E$, (b) duration $\tau$, (c) peak power $P$, (d) wait time $w$, for a case with $\eta = \nu = 0.005$ using a resolution of $64^2 \times 16$.

From a scatter plot (not shown), a power law $P\tau \sim E^{1.034 \pm 0.001}$ was found to be extremely well fit of the data just as the first case shown above. Using $\alpha_P = 2.52$, $\alpha_\tau = 3.13$, and $\alpha_E = 1.97$ observed from Fig. 4, we get $\gamma = 1.09$ from Eq. (3), which is consistent with this relation. Similar self-consistent tests were applied to the other two cases with good agreements. The values of $\alpha_E$, $\alpha_\tau$, $\alpha_P$, and $\alpha_w$ are then plotted for the four cases in Fig. 5 versus the resistivity $\eta$. We see that the values for the two cases with smaller $\eta$ are similar, suggesting a saturation, while the values for the two cases with larger exhibit more variation. From scatter plots (not shown), there is no significant correlation between the wait time $w$ and the other quantities. However, from the geometrical meaning of the event identification process, it is expected that the wait time distribution should have a statistics similar to the duration distribution, since $w$ can be regarded as the duration of events with negative $W - \overline{W}$. We see that indeed the values of $\alpha_w$ are similar to those of $\alpha_\tau$, especially in the small $\eta$ limit.

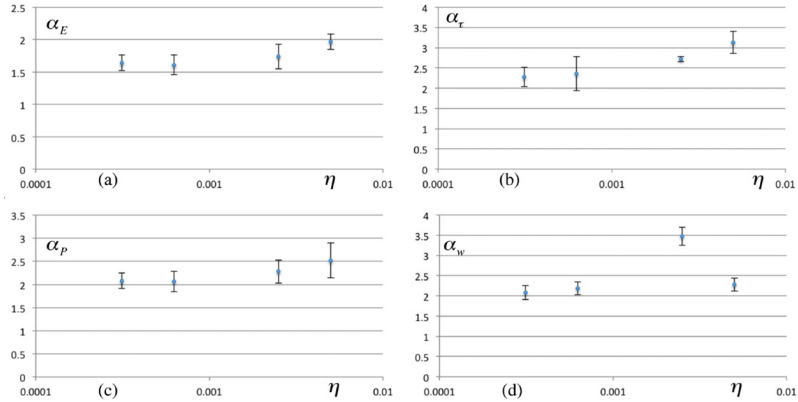

**FIGURE 5.** Plot of (a) $\alpha_E$, (b) $\alpha_\tau$, (c) $\alpha_P$, (d) $\alpha_W$, versus resistivity $\eta$ (= $\nu$) used in four runs.

# DISCUSSION AND CONCLUSION

We have presented a statistical analysis of heating events (nanoflares) based on direct 3D RMHD simulations. Our results, especially the values of $\alpha_E$ are similar to those found by using 2D MHD simulations [8,9] under the same method of analysis. The values of the four power law indices found here are also within the range of values based on solar flares observations, based on comparison with Refs. [5,6,11] and other references not cited here. The fact that $\alpha_E$ found are consistently less than 2 is somewhat in contradiction to the importance of heating based on nanoflares. However, we caution that our analysis is still based on the total heating rate over the whole volume which usually contains a number of current sheets, as shown in Fig. 1(a). It could be argued that it is a single current sheet that should be regarded as a nanoflare. Distributions based on identification of current sheets are of course more difficult and are left for future studies.

# ACKNOWLEDGMENTS

This work is supported by NASA grants NNX08BA71G, NNX06AC19G, a NSF grant AGS-0962477, and a DOE grant DE-FG02-07ER54832.